\documentclass[preprint,12pt]{elsarticle}

\usepackage{graphics}
\usepackage[percent]{overpic}

\journal{Colloids Surface A}

\begin{document}

\begin{frontmatter}

\title{Reconsideration on structural anisotropy of silica hydrogels prepared in magnetic field}
\author{Atsushi Mori$^{a,1}$, Takamasa Kaito$^{b,2}$, Hidemitsu Furukawa$^{c}$}

\address{$^a$Institute of Technology and Science, Tokushima University \\
$^b$Graduate School of Engineering, Tokushima University \\
$^c$Graduate School of Science and Engineering, Yamagata University \\[1.5ex]
$^1$Corresponding author; Postal address: Institute of Technology and Science, Tokushima University, Tokushima 770-8079, Japan; Tel: +81-88-656-9417, Fax: +81-88-656-9435, E-mail: atsushimori@tokushima-u.ac.jp \\
$^2$Present address: KRI Inc.}
%
%\sloppy
%
%\maketitle

\begin{abstract}
In a previous paper [A. Mori, T. Kaito, H. Furukawa, Mater. Lett. 62 (2008) 3459-3461], we carried out birefringence measurements of Pb(II)-doped silica hydrogels prepared in a magnetic field ($\mbox{\boldmath{$B$}}$).
For a 5T sample, we observed a negative birefringence with the optic axis along $\mbox{\boldmath{$B$}}$.
At that time, providing a positive intrinsic birefringence of silica, we speculated that in the birefringent gels the gel network extended perpendicular to $\mbox{\boldmath{$B$}}$.
The purpose of this paper is to reconsider this speculation on the basis of previous and recent results [T. Kaito, S.-i. Yanagiya, A. Mori, M. Kurumada, C. Kaito, T. Inoue, J. Cryst. Growth 289 (2006) 275-277; T. Kaito, A. Mori, C. Kaito, J. Chem. Chem. Eng., 9 (2015) 61-66].
In the former, the silica gels were used as a medium of a crystal growth of PbBr${}_2$ and aligned arrays of crystallites with long axis parallel to $\mbox{\boldmath{$B$}}$ were obtained.
In the latter, Pb(II) nanocrystallites were formed in silica xerogels by electron irradiation.
Both of the short axis of PbBr${}_2$ crystallites and the diameter of Pb(II) crystallites were a few tens of nanometers.
This size must be a size of short axis of pores in the gel networks elongated affected by the magnetic field.
Since the PbBr${}_2$ crystallites were elongated along the magnetic field, we conclude that the Pb(II)-doped silica gel networks aligned along the magnetic field.
\end{abstract}


\date{June 25, 2015}

\end{frontmatter}

\section{Introduction}
We grew PbBr${}_2$ crystals using silica hydrogels prepared in a magnetic field of $B$ = 5 T as media of crystal growths \cite{Kaito2006JCG289,Kaito2006JCG294}.
Aligned arrays of nanocrystallites are found in gels \cite{Kaito2006JCG289}.
The magnetic field was applied during the preparation of the gels and the crystallographic axes of the crystallites were oriented along the direction of the magnetic field. 
Even if a magnetic field was applied during the crystal growth, it did not affect appreciably.
Thus, it is anticipated that the magnetic field brought a structural anisotropy in the Pb(II)-doped silica hydorgels during the preparation.
Identification of structural anisotropy has, so far, been one of our recent subjects.

There are a lot of potential uses of the silica gels with controlled structure.
Therefore, control of the structure of silica gels has, so far, been studied.
The transport of materials in the gels depends on the structure.
Aerogels can be used as media in column chromatography.
If pore size is highly controlled uniformly, the filtering of the column must be improved.
One can intuitively realize anisotropy in mechanical and/or thermomechanical properties due to the anisotropic structure.
As described in a monograph \cite{Ilre} and reported in literatures such as Refs.~\cite{Takahashi1995,Birch2000,Knoblich2001,Hoffmann2006,Rahman2015}, one can control the structure of silica gels by the selection of starting materials, pH control of the solvent, solvent exchange during polymerization stage, and aging.
In addition, porosity of commercially available silica gels for column chromatography was modified by several method \cite{Leboda1995_181,Leboda1995_191,Goworek1999,Leboda2000,Gunko2004}.

Magnetic field applied during the gelation affects the network structure of polymer gels.
Chemically cross-linked poly($N$-isopropylacrylamid) forms aligned network structure perpendicular to the magnetic field \cite{Otsuki2006}.
Physically cross-linked agarose gels also exhibit the perpendicular alignment \cite{Yamamoto2006}.
If side-chain groups prefer the parallel alignment, main chains align perpendicularly.
On the other hand, in a case that a magnetic moment is included in a group of the main chain, the main chains align parallel \cite{Shigekura2005}.

Because silica is diamagnetic, one cannot imagine the alignment of silica polymers due to direct interaction with a magnetic field.
In a previous paper \cite{Mori2008}, basically base on the results of birefringence measurement and the discussions on a likely mechanism of interaction with the magnetic field, we reached incorrectly at a conclusion that the silica gel network must extend perpendicular to the magnetic field.
A negative birefringence on the order of $10^{-6}$ for 5T samples, which was a result from a devised S\`{e}narmont method as described in Ref.~\cite{Mori2008} (also, see Ref.~\cite{Mori2014} for corrections of typographical errors in Ref.~\cite{Mori2008}), was interpreted as follows.
There have been an interest in the structure of scale around several tens or a few hundreds of nanometers.
Pores of such scale have been reported in silica aerogels \cite{Wang1998} and commercial gels for column chromatography \cite{Goworek1999,Stefaniak2003}.
Also, along with fractal nature, which were more often reported for silica xerogels (such as in Refs.~\cite{Takahashi1995,Birch2000,Knoblich2001,Hoffmann2006}), such scales were observed in silica hydrogels \cite{Schaefer1984,Ferri1991}.
It is natural to imagine the existence of closed loops of the same scale in the hydrogels.
Let us remember the presence of Pb$^{++}$ ions and that the skeletons made of silicon and oxygen atoms possess a lot of dangling bonds in hydrogels.
We speculate ring currents along close loops due to complexes formed on such skeletons with Pb$^{++}$ ions through a mechanism similar to that of electric conduction of the conjugated polymers \cite{Heeger1988} and the force tending to direct those rings perpendicular to the magnetic field.
In this way, providing a positive intrinsic birefringence of silica, in Ref.~\cite{Mori2008} we struggled to give an interpretation to the negative birefringence from this speculated structural anisotropy.
Narrowing of the pore size distribution, which resulted from the scanning microscopic light scattering (SMILS) \cite{Furukawa2003} performed in Ref.~\cite{Mori2008}, can be understood consistently, too.
It should be noted that, irrespective of the direction of magnetic alignment, certain characteristic size distributes narrowly if ordering occurs.

The aim of this paper is to overturn the conclusion of Ref.~\cite{Mori2008} on the structural anisotropy in silica hydrogels prepared in a high magnetic field.
To do so, a consideration is given relying on a previous result \cite{Kaito2006JCG289} and a recent one \cite{Kaito2015} with help of a result of the reinvestigated birefringence \cite{Mori2014}.
Further analysis of SMILS result in Ref.~\cite{Mori2008} is also given.

\section{Materials and Methods}
In Ref.~\cite{Mori2008} as well as in Refs.~\cite{Kaito2006JCG289,Tomita2010}, the samples were prepared in the same way as described previously \cite{Kusumoto2005}, except for the application of magnetic field.
The starting material was sodium metasilicate, the acid to maintain the solution strongly acidic was acetic acid, and the source of Pb(II) was a Pb(NO${}_3$)${}_2$ aqueous solution.

In Ref.~\cite{Kaito2015}, unlike previous studies \cite{Kaito2006JCG289,Kaito2006JCG294,Mori2008}, a lead (II) acetate aqueous solution was employed as the source of Pb(II).
After preparing silica hydrogels in a magnetic field of various strengths like in Refs.~\cite{Kaito2006JCG289,Kaito2006JCG294,Mori2008}, the samples were dried in test tubes as prepared for a year.
Then, electrons were irradiated to the samples in a transmission electron microscope (TEM) environment.
We observed the samples by a TEM (Hitachi H-9000NAR).

We should note on the difference between the mechanisms for crystallization in those two kinds of experiments.
In the former, crystallization is governed by the diffusion of Br$^{-}$ ions and reaction with Pb$^{++}$ ions thereafter.
In the latter, such diffusion-reaction mechanism does not exist.
Pb$^{++}$ ions are reduced by irradiated electrons and then metallic Pb clusters precipitate.

\section{Results and Discussion}
\begin{figure}[htb]
\begin{center}
\includegraphics[width=0.9\textwidth, keepaspectratio]{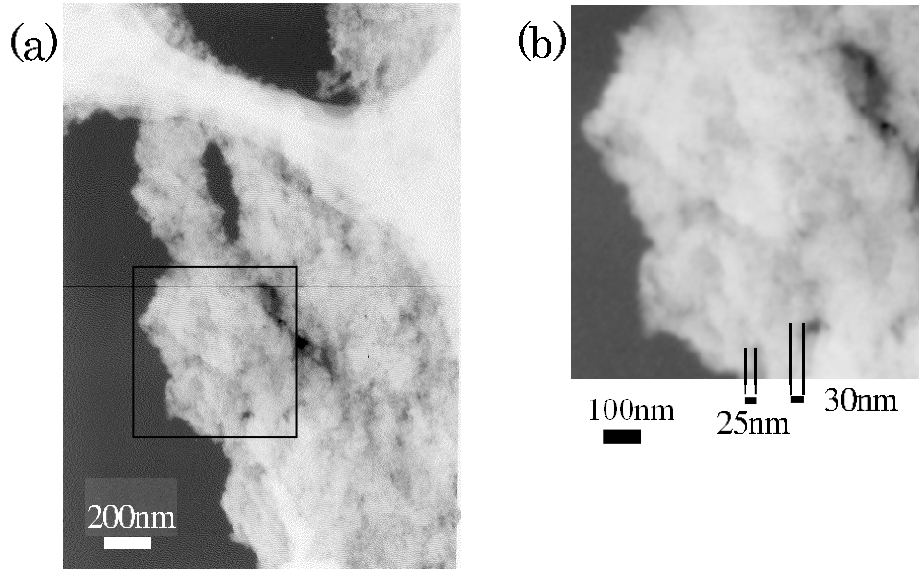}
\end{center}
\caption{\label{fig:EI} (a) A TEM image of Pb(II) nanocrystallites formed inside macropores of a silica xerogel prepared in 10 T induced by electron irradiation \cite{Kaito2015} (this image is not a reuse of one in Ref.~\cite{Kaito2015}), and (b) a magnification of a part of (a), indicated by a square.
Sharpness has been reduced to recover the quality of magnified image.
Among many crystallites two pairs of vertical lines have been added to two of them for guide for eyes in (b).}
\end{figure}

\begin{figure}[htb]
\begin{center}
\includegraphics[width=0.95\textwidth, keepaspectratio]{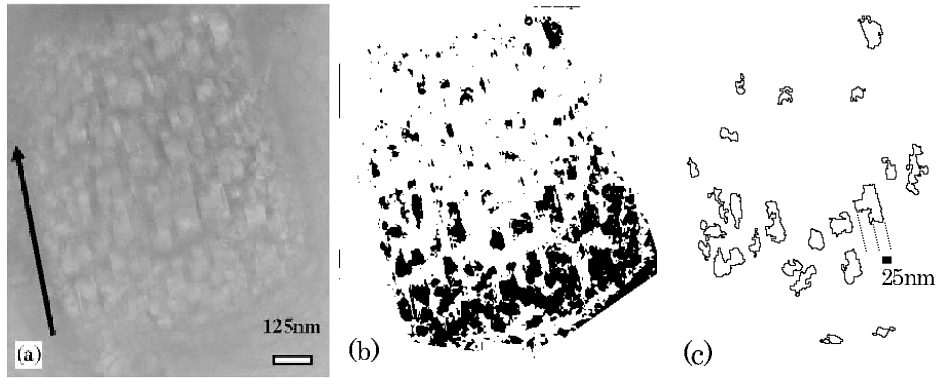}
\end{center}
\caption{\label{fig:JCG} (a) A TEM image of an aligned array of PbBr$_2$ nanocrystallites formed in a silica hydrogel [Fig.~3(a) of Ref.~\cite{Kaito2006JCG289}], (b) white-black binary image of (a), and (c) the particles pucked up by particle analysis of ImageJ software.
Auxiliary lines are drawn beneath one of merged objects in (c).
The direction of magnetic field is indicated by the arrow in (a).}
\end{figure}

We have recently reported formation of Pb(II) nanocrystallites induced by electron irradiation inside macropores of Pb(II)-doped silica xerogels prepared in a magnetic field of various strengths \cite{Kaito2015}.
A TEM image is shown in Fig.~\ref{fig:EI}.
Nanocrystallites are observed in aggregates inside macropores.
A part of Fig.~\ref{fig:EI}(a) is taken up in Fig.~\ref{fig:EI}(b).
Sharpness has been reduced and two pairs of vertical lines have been inserted auxiliary.
Their sizes are a few tens of nanometers, which coincide to the length of short axis of PbBr${}_2$ nanocrystallites in Fig.~3(a) of Ref.~\cite{Kaito2006JCG289} [carried into this paper as Fig.~\ref{fig:JCG}(a)].
It is noted that due to the reduction of sharpness we failed to put smaller crystallites in relief as indicated.
Those sizes must be the length of short axes of pores elongated affected by the magnetic field in the hydrogels.
We guess that similar structures remain inside macropores in the xerogels.
During drying process of hydrogels, macropores form.
Therefore, the microstructure of xerogels also more or less differs from that of hydrogels.
Nevertheless, there must remain gel network structure in macropores similar to that of hydrogels.
From the fact that the orders are the same, we speculate that the length of the short axis of pores in the hydrogels is also on the same order.

From the same point of view, the long axes of PbBr${}_2$ nanocrytallites must correspond to the long axes of pores.
In Fig.~\ref{fig:JCG}(a), nanocrystallites are elongated in the direction of magnetic field.
Remember the difference in the mechanisms of crystallizations as pointed out in the preceding section.
In the crystal growth in hydrogels used as media, diffusion controls the growth.
Moreover, nutrients are supplied by diffusion in hydrogels.
Therefore, crystallites can grown along the elongated micropores.
On the hand, in crystallization in xerogels induced by an electron irradiation diffusion of nutrient does not govern the crystal growth.
Unless coalescence occurs, the precipitates may grow in a spherical-like form.
It is speculated that order of the underlaying gel networks or that of self-similar network structure whose scale is less than that affects to the outer form of the precipitates must influences the crystallinity of precipitates (discussion with regard to the fractal nature of the gel network on micropore scale will be given later).
It may result in faceting behavior; however, Fig.~\ref{fig:EI} does not possess a sufficient resolution.
We do not wish to improve the resolution because it is not the present conc1ern.
A rough estimation is possible from Fig.~\ref{fig:JCG}(a).
Dimension of the nanocrystallites was evaluated using ImageJ software.
After making the TEM image [Fig.~\ref{fig:JCG}(a)] into a white-black binary image, particle analysis was carried out.
A binarized image is shown in Fig.~\ref{fig:JCG}(b) and a resultant drawing of particle analysis in Fig.~\ref{fig:JCG}(c).
Result of binarization depends on the threshold, and thus the result of particle analysis is not very unique.
The average lengths of major and minor axes were 80 and 40 nm in approximation by ellipses.
Regarding the present TEM image approximation by rectangles brought vague results due to merged objects made of a number of particles.
Dotted lines are drawn to one of merged ones for guide, from which one can understand a tendency of overestimation even by the approximation by ellipses.
One can separate two particles contacting each other by selecting the threshold.
With the threshold which can diminish the border of objects and separate merged ones, the size of objects decreases.
This situation occurs in general.
Along with the fact that the pore in the network only give the upper limit of the precipitate dimension, the present estimation is underestimated for the pore dimension.

\begin{figure}[htb]
\begin{center}
\includegraphics[width=0.78\textwidth, keepaspectratio]{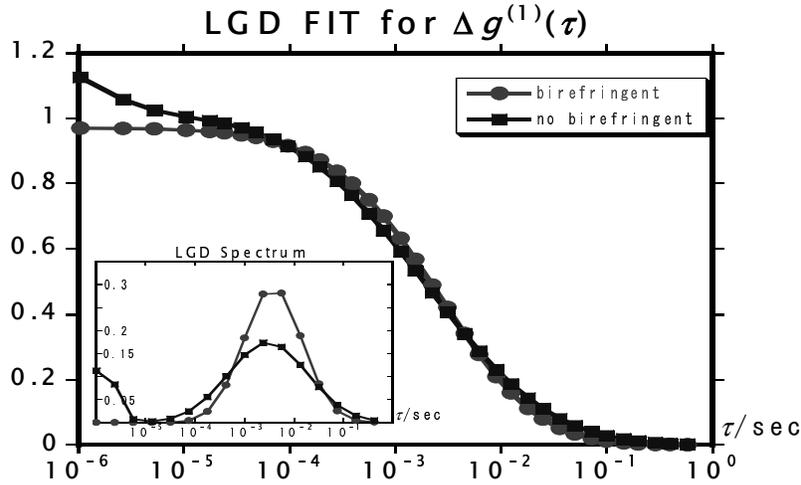}
\end{center}
\caption{\label{fig:ML} A result of SMILS for the scattering angle of $60^{\circ}$ (Fig.~4 of Ref.~\cite{Mori2008}).
In the inset peaks of relaxation time distributions locate in the range of millisecond for both birefringent and no-birefringent samples.
The corresponding characteristic length is on the order of hundreds nanometers.}
\end{figure}

A rough estimation for a previous SMILS experiment \cite{Mori2008} on a characteristic length gives a support.
Figure~4 of Ref.~\cite{Mori2008} is carried as Fig.~\ref{fig:ML}.
The inset is a logarithmic Gaussian distribution spectra of the relaxation times ($\tau$) for autocorrelation functions of the electric field of scattered light.
Peaks appear in the rage of millisecond for both samples.
The diffusion coefficient $D$ of the corrective diffusion mode is usually estimated by fitting by $\tau^{-1} = Dq^2$, where $q$ denotes $|\mbox{\boldmath{$q$}}|$ with $\mbox{\boldmath{$q$}}$ being the scattering vector.
Since the working wavelength in this measurement was $\lambda$ = 532 nm, $q$ = $(4\pi n/\lambda)\sin(\theta/2)$ with $\theta$ being the scattering angle is evaluated as $1.57\times 10^{7}$ m$^{-1}$ for $\theta$ = $60^{\circ}$ \cite{theta} ($n$ is the diffactive index of the media).
Therefore, for the result of Fig.~\ref{fig:ML} one can estimate $D$ on the order of  $10^{-12}$ m$^2$/sec.
Owing to the Stokes-Einstein relation $D = k_BT/6\pi\eta\xi$ the corresponding characteristic length $\xi$ is roughly estimated as one hundred and several tens of nanometers or a few hundreds of nanometers, where $k_BT$ is the temperature multiplied by Boltzmann's constant and $\eta$ the viscosity coefficient of the solvent. 
Taking into consideration a note that the estimate of 80 nm was underestimated, this characteristic length is speculated to correspond to a long axis of elongated pores.

The present concern has been limited to the direction of magnetic alignment in a scale of micropore.
Pore size distribution has not been determined.
If there is a fractal nature in the scale under consideration, to determine the distribution around a mean value is meaningless because of a scale-free property.
The dimension of nanocrystallites was not limited only by the compartment within the micropores.
Even if one can determine the nanocrystallite size distribution, it unnecessarily coincide to the pore size distribution.

We conclude that the silica gel networks align along the magnetic field.
This conclusion is opposite to that of Ref.~\cite{Mori2008}.
This is , however, consistent with a recent result on the sign of measure of birefringence of silica hydrogels prepared in a magnetic field of various strengths \cite{Mori2014}.
We have also overturned a conclusion on the sign of birefringence of Ref.~\cite{Mori2008}.
In Ref.~\cite{Mori2008}, the vessels used as sample cells were cylindrical tube made of borosilicate glass.
In Ref.~\cite{Tomita2010}, we performed the same measurements using square cross-sectional cells made of both borosilicate glass and quartz.
Special cares were payed to selection of cells so that the optical path difference of empty cells was on the order of $10^{-1}$nm.
Nevertheless, obtained results were qualitatively different from each other.
The former indicted decreasing property of $\Delta n (B)$ and negative birefringence was measured for $B$ $>$ 5 T.
A negative birefringence on the order of $10^{-6}$ was not reproduced for 5T samples.
Instead, negative birefringence on the order of $10^{-7}$ was measured for 7T and 10T samples.
However, we infer that the resultant negative birefringence was attributed to the effect of the borosilicate glass cell wall \cite{Mori2014}.
It is known that borosilicate glasses include a lot of impurities, and thus the cell wall surfaces are not featureless.
Therefore, alignment of silica chains is affected by the surface structure and/or other unexpected factors due to impurities.
On the other hand, a trend of positive birefringence was seen in the latter.
Effect of cell wall on the alignment of silica chains is expected to be smaller for quartz glass than for borosilicate glass.
In Ref.~\cite{Mori2014}, we have reinvestigated the results for quartz cells and overturned the conclusion of Ref.~\cite{Mori2008} on the sign of birefringence.
That is, the silica hydrogels prepared in a high magnetic field have a positive birefringence with the optic axis along the direction of magnetic field.
Providing a positive intrinsic birefringence, one reaches to a conclusion that silica chains tend to align parallel to the magnetic field

Let us discuss the mechanism of magnetic alignment of the silica network in hydrogels.
In Ref.~\cite{Mori2008}, we assumed excitation of soliton type, such as for conducting polymers \cite{Heeger1988}, on silica chain with presence on Pb${}^{++}$ ions and this excitation plays a role of electric conduction carrier.
Ring currents on loops in gel networks make magnetic moments perpendicular to the loops.
Accordingly, the loops tend to align perpendicular to the magnetic field.
Now, we point that excitation is not limited to such type; excitation accompanying magnetic moments can arise.
In such case, silica chains align parallel to the magnetic field. 
Confirmation of this speculation is a subject of future researches.

\section{Concluding remarks}
We wish to conclude this paper with the following sentence.
A conclusion of a previous paper \cite{Mori2008} has been overturned: a silica gel network in Pb(II)-doped silica hydrogels aligns parallel to a magnetic field.

Let us close this paper with a comment on a future research.
We have constructed a set up of the S\`{e}narmont method with a mechanical rotating analyzer to reduce the measurement time \cite{Mori2015}.
In previous studies~\cite{Mori2008,Tomita2010}, the numbers of measurement were small because the measurements were extremely time-consuming.
We expect an improvement of statistics accuracy to support the conclusion on the sign, and to update the magnitude of the measure of birefringence of silica gels.

\section*{References and notes}


\begin{thebibliography}{99}
\bibitem{theta} The scattering angle for Fig.~3 of Ref.~\cite{Mori2008} should be read as 60$^{\circ}$. This does not affect the related conclusion therein.
\end{thebibliography}
\end{document}